\documentclass[prl,twocolumn,showpacs,preprintnumbers,amsmath,amssymb]
{revtex4}

\usepackage{graphicx}
\usepackage{amsmath}
\usepackage{amsbsy}
\usepackage{dcolumn}
\usepackage{bm}

\begin{document}

\title{Spin dynamics in the stripe phase of the cuprates}

\author{Brian M\o ller Andersen and Per Hedeg\aa rd}

\affiliation{\O rsted Laboratory, Niels Bohr Institute,
Universitetsparken 5, 2100 Copenhagen \O, Denmark}

\date{\today}

\begin{abstract}
Within a model that supports stripe spin and charge
order coexisting with a d$_{x^2-y^2}$-wave superconducting phase, we study
the self-consistently obtained electronic structure and the associated transverse dynamical spin
susceptibility. In the coexisting phase of superconducting and static
stripe order, the resulting particle-hole continuum can strongly damp
parts of the low-energy spin wave branches. This provides insight into recent inelastic neutron
scattering data revealing the dispersion of the low-energy collective
magnetic modes of lanthanum based cuprate superconductors. 
\end{abstract}


\maketitle

The electronic properties of the underdoped cuprates is dominated by competing instabilities and
coexistence of several ordered states.
An example is given in the accumulating evidence the doped holes tend to
self-organize into one-dimensional rivers of
charge\cite{originalstripesnesting}. 

For instance, in Nd-doped La$_{2-x}$Sr$_x$CuO$_4$ neutron scattering (NS)
revealed a quartet of incommensurate (IC) elastic magnetic peaks at $((1 \pm \delta)\pi,\pi)$,
$(\pi,(1 \pm \delta)\pi)$ and Bragg charge peaks at $(\pm 2\pi\delta,0)$,
$(0,\pm 2\pi\delta)$\cite{tranquada}. This is consistent with the doped holes
forming 1D domain walls oriented along the Cu-O bonds in the CuO$_2$
planes. These stripes are separated by $1/\delta$ (in units of
the lattice spacing $a$) and the staggered 
magnetization gains an extra phase shift $\pi$ when crossing a
stripe. Later, it was realized that d-wave superconductivity (dSC) coexists with the
static stripe order\cite{tranquadasc}. The NS results from pure La$_{2-x}$Sr$_x$CuO$_4$
exhibits similar IC peaks\cite{cheong}. For these materials, Bragg peaks are observed for
$0.02<x<0.13$, whereas for $x>0.13$ a small doping dependent spin gap
opens in the magnetic excitation spectrum\cite{wakimoto}. 

The dispersion of the IC peaks, i.e. $\delta(\omega)$, between 0-40
meV in optimally and underdoped LSCO was  
measured recently by Christensen {\sl et al.}\cite{niels}. It
was found that both in the pseudogap and dSC state, the IC peaks disperse
toward $(\pi,\pi)$ as the energy
increases but with no sign of an intense resonance feature at 
$(\pi,\pi)$. Furthermore, the IC peaks broaden as the energy
increases. A similar dispersion was found in La$_{2-x}$Ba$_x$CuO$_4$
by Tranquada {\sl et al.}\cite{tranqLBCO}. In La$_{2-x}$Ba$_x$CuO$_4$
and YBa$_2$Cu$_3$O$_{6+x}$ it was further found that above the $(\pi,\pi)$ crossing, the spin
response is dominated by four peaks rotated $\pi/4$ relative to the
low-energy IC peaks\cite{tranqLBCO,hayden}. These high-energy peaks disperse to
larger wave vectors with increasing energy transfer. In YBCO the low-energy
spin response is dominated by the commensurate (C) $(\pi,\pi)$ resonance which
disperses downward with decreasing energy\cite{araimook}. This points to a degree of ubiquity
in the spin fluctuation spectrum of the cuprates. The main difference
between the materials appears to be the size of the doping dependent spin gap and
the intensity distribution along the spin branches. These experiments
have sparked new theoretical studies dealing mostly with the high-energy response\cite{newtheo}.

In this Letter we report the self-consistent results of the electronic structure and
the corresponding transverse spin susceptibility within a model that
supports static IC spin and charge density wave
solutions. In particular, we focus on the low-energy spin dynamics and
the influence of the dSC on the intensity distribution of the spin branches when it coexists
with static spin and charge order. These studies are motivated by the
above-mentioned new experimental insight, and the strong evidence for
stripes in the lanthanum based materials. A similar approach was used to study the phonon anomalies caused by
collective modes in the stripe phase of the 2D Hubbard model without
allowing for dSC order\cite{kaneshita}.

Several previous studies of magnetic modes in
d-wave superconductors have started from {\sl spatially homogeneous}
phases\cite{levin,bulutnorman}. In contrast, recent 
spin-only models have also been proposed to describe the spin dynamics in the IC stripe 
phase\cite{batista}. Here, we bridge these two apparently complementary approaches by solving
self-consistently the following minimal model defined 
on a 2D lattice 
\begin{eqnarray}\label{hamiltonian}\nonumber
\hat{H}= &-& \sum_{\langle ij \rangle\sigma}  \left( t_{ij}
\hat{c}_{i\sigma}^{\dagger}\hat{c}_{j\sigma}  +
\mbox{H.c.} \right) + \sum_{i\sigma} \left( U
\langle \hat{n}_{i\overline{\sigma}} \rangle - \mu \right)
\hat{n}_{i\sigma} \\ 
&+& \sum_{\langle ij \rangle} \left( \Delta_{ij}
\hat{c}_{i\uparrow}^{\dagger}\hat{c}_{j\downarrow}^{\dagger} +
\mbox{H.c.} \right).
\end{eqnarray}
Here, $\hat{c}_{i\sigma}^{\dagger}$ creates an
electron of spin $\sigma$ on site $i$, $t_{ij}=t,t'$ denote the
first and second nearest neighbor hopping integrals, $\mu$ is the chemical
potential, and
$\hat{n}_{i\sigma}=\hat{c}_{i\sigma}^{\dagger}\hat{c}_{i\sigma}$ is
the occupation number on site $i$.
The model (\ref{hamiltonian}) is the mean-field version of the extended Hubbard model
with onsite repulsion $U$ and nearest neighbor attraction $V$ and is
aimed to mimic essential features of phases of coexisting spin, charge and dSC
order. The nearest neighbor attraction $V$ triggers the singlet dSC at
the mean-field level, $\Delta_{ij}=V \left( 
  \langle \hat{c}_{i\uparrow}\hat{c}_{j\downarrow} \rangle - \langle
  \hat{c}_{i\downarrow}\hat{c}_{j\uparrow} \rangle
  \right)$. This approach has previously been used extensively to gain
insight into the electronic structure in phases of coexisting order\cite{previous,ichioka}.

As is well-known, away from half-filling the Hamiltonian (\ref{hamiltonian})
produces inhomogeneous spin and charge order $S^z_i=\frac{1}{2} (\langle \hat{n}_{i\uparrow}
\rangle - \langle \hat{n}_{i\downarrow} \rangle)$ and $\rho_i=(\langle \hat{n}_{i\uparrow}
\rangle + \langle \hat{n}_{i\downarrow} \rangle)$ with the modulation
period of $S^z_i$ being exactly twice the period of $\rho_i$. This is the stripe phase at the
mean-field level. In general, i.e. when $t' \neq 0$, the stripes
are metallic since the mid-gap states cross the Fermi
level\cite{ichioka}.

For an ordered array of stripes with spin periodicity $N$ we can divide the real-space lattice
into smaller supercells of size $N$. Then, the Hamiltonian can be diagonalized by
the Bogoliubov transformation,
${\hat{c}_{i\sigma}}^\dagger = \sum_{n{\mathbf{k}}}
(u^*_{n{\mathbf{k}}\sigma}({\mathbf{r}}_i) e^{-i {\mathbf{k}} \cdot {\mathbf{R}}_i}
\hat{\gamma}_{n{\mathbf{k}}\sigma}^\dagger 
+ \sigma v_{n{\mathbf{k}}\overline{\sigma}}({\mathbf{r}}_i) e^{i {\mathbf{k}} \cdot {\mathbf{R}}_i}
{\hat{\gamma}}_{n{\mathbf{k}}\overline{\sigma}})$,
where ${\mathbf{r}}_i$ denotes a site within the supercell which in turn is positioned at
${\mathbf{R}}_i$.
The wave vectors ${\mathbf{k}}$ belong to the corresponding reduced
Brillouin zone, and $\sigma =
+(-)1$ for up(down) spin. This results in a set of Bogoliubov-de Gennes equations
to be diagonalized for each ${\mathbf{k}}$. The self-consistency is enforced through
iteration of the relations,
$1-n_h = \frac{1}{N}\sum_{i\sigma} \langle \hat{n}_{i\sigma} \rangle$,
$n_h$ is the hole doping, and
\begin{eqnarray}
\langle \hat{n}_{i\sigma}\! \rangle\!\!\!&=&\!\!\!\! \sum_{n{\mathbf{k}}}\! \left[
  |u_{n{\mathbf{k}}\sigma}(i)|^2 f(E_{n{\mathbf{k}}\sigma}) 
\!+\! |v_{n{\mathbf{k}}\overline{\sigma}}(i)|^2
f(-E_{n{\mathbf{k}}\overline{\sigma}})\right],\\
\Delta_{ij}\!\!\!&=&\!\!\!\sum_{n{\mathbf{k}}} \left[
  v_{n{\mathbf{k}}\sigma}^*(i) u_{n{\mathbf{k}}\sigma}(j)
  f(E_{n{\mathbf{k}}\sigma}) \right. \\\nonumber
&-& \left. u_{n{\mathbf{k}}\sigma}(i) v_{n{\mathbf{k}}\sigma}^*(j)
f(-E_{n{\mathbf{k}}\sigma})\right].
\end{eqnarray}
As usual, $f(E)=[1+\exp(E\beta)]^{-1}$ denotes the Fermi distribution
function with $\beta=1/kT$.

Below, we report the spin dynamics obtained from the stable
configurations both with and without dSC when $U=4.0t$, $n_h=0.125$
and $N=8$. In principle, for a given set of parameters $U$, $V$, $t'$, 
and $n_h$ the system may prefer a spin period different from $N=8$. By
varying $n_h$ we have checked that this possibility does not 
qualitatively alter the results below.    

\begin{figure}[t]
\includegraphics[width=8.5cm]{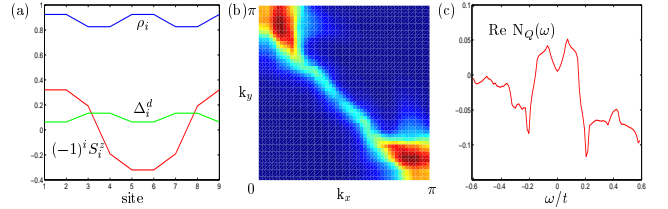}
\caption{\label{configs}(a) Bond-centered spin density $(-1)^{i_x} S^z_i$
  (red), charge density $\rho_i$ (blue) and pairing potential $\Delta^d_i$
  (green) versus site obtained when $U=4.0t$, $V=2.0t$, 
  $t'=-0.37t$, (b) the spectral
  weight $I({\mathbf{k}})$ ($\Delta \omega=0.1t$) versus ${\mathbf{k}}$
  for disordered stripes. (c) plot of
  the real part of $N_{\mathbf{Q^*}}(\omega)$ versus energy $\omega$.}
\end{figure}

A typical example of the self-consistent results for the spin
(charge) density  $S^z_i$ ($\rho_i$)
and the pairing potential $\Delta^d_i=(\Delta_{i,i+e_x} + \Delta_{i,i-e_x} -
\Delta_{i,i+e_y} - \Delta_{i,i-e_y})/4$ is shown in
Fig. \ref{configs}a. Clearly, this shows the expected anti-phase
stripe ordering with concomitant modulations of $\Delta^d_i$. The
bond-centered solutions are found to have energies slightly lower than
the site-centered stripes.

Turning briefly to the electronic structure obtained from these 
self-consistent solutions, it is well-known that for the spectral
weight, $I({\mathbf{k}})= \int_{\mu-\Delta w}^{\mu}
A({\mathbf{k}},\omega) d\omega$, the stripes generate weight around
the antinodal regions\cite{ichioka,salkola}. 
Here $A({\mathbf{k}},\omega)$
is the single-particle spectral function and $\Delta w$ is an integration window below the chemical
potential $\mu$. For instance, in Fig. \ref{configs}b we show
$I({\mathbf{k}})$ for $0 \leq k_x,k_y \leq \pi$ for the same
parameters used in Fig. \ref{configs}a but allowing for stripe
disorder\cite{footnote1}. This figure is strikingly similar to the
recent ARPES data by Zhou {\sl et al.}\cite{zhou} 
Furthermore, the stripe ordering causes a modulation of the pairing potential as
seen in Fig. \ref{configs}a which influences the Fourier
transform of the LDOS $N_{\mathbf{q}}(\omega)$\cite{ichioka,podolsky}. For
instance, as seen from Fig. \ref{configs}c, the real part of
$N_{\mathbf{Q^*}}(\omega)$ versus energy 
$\omega$ at the charge ordering vector
${\mathbf{Q^*}}=(\frac{3\pi}{2},0)$ exhibits two low-energy zero crossings in
agreement with recent STM measurements by Howald {\sl et al.}\cite{howald}.

Motivated by the agreement of the self-consistent mean-field solutions
of Eqn. (\ref{hamiltonian}) and the mentioned electronic probes, we turn now to our
main topic: the collective spin dynamics in the 
stripe phase of Eqn. (\ref{hamiltonian}). The stripes explicitly break the SU(2) spin rotation
symmetry and hence associated Goldstone modes are expected in the transverse spin
susceptibility $\chi^{+-}({\mathbf{q}},\omega)$. Below, we investigate the effect of  
the charges and the dSC order on the dynamic part of the spin modes. An explicit
calculation of the Fourier transform of $\chi^{+-}({\mathbf{r}}_i,{\mathbf{r}}_j,\tau) = - \langle 
T_\tau \hat{S}^+_{j}(\tau)\hat{S}_{i}^-(0)
\rangle$ shows that to Gaussian order the response is given by the
diagonal elements of the $N \times N$ matrix
\begin{eqnarray}\label{fullsus}
\chi^{+-}({\mathbf{q}},\omega)=\chi^{+-}_0({\mathbf{q}},\omega) \left(1-U
\chi^{+-}_0({\mathbf{q}},\omega)\right)^{-1}.
\end{eqnarray}
Here the matrix elements of the bare susceptibility
$\chi^{+-}_0({\mathbf{q}},\omega)_{QQ'}$ are given by 
\begin{widetext}
\begin{eqnarray}\label{monster}
\chi^{+-}_0({\mathbf{q}},\omega)_{QQ'} &=& \frac{1}{4N^2}
\sum_{\stackrel{\scriptstyle {\mathbf{k}}nm}{{\mathbf{r}}_i{\mathbf{r}}_j\sigma}} \left[
a_1(u,v)
\frac{1-f(E_{n{\mathbf{k}}\sigma})-f(E_{m{\mathbf{k}}+{\mathbf{q}}\sigma})}{\omega
+ E_{m{\mathbf{k}}+{\mathbf{q}}\sigma} + E_{n{\mathbf{k}}\sigma} + i\Gamma} +
a_2(u,v)
\frac{f(E_{n{\mathbf{k}}\sigma})+f(E_{m{\mathbf{k}}+{\mathbf{q}}\sigma})-1}{\omega
- E_{m{\mathbf{k}}+{\mathbf{q}}\sigma} - E_{n{\mathbf{k}}\sigma} + i\Gamma} \right.
\\\nonumber &+&
\left. b_1(u,v)
\frac{f(E_{n{\mathbf{k}}\sigma})-f(E_{m{\mathbf{k}}+{\mathbf{q}}\overline{\sigma}})}{\omega
+ E_{m{\mathbf{k}}+{\mathbf{q}}\overline{\sigma}} -
E_{n{\mathbf{k}}\sigma} + i\Gamma} +
b_2(u,v)
\frac{f(E_{m{\mathbf{k}}+{\mathbf{q}}\sigma})-f(E_{n{\mathbf{k}}\overline{\sigma}})}{\omega
+ E_{n{\mathbf{k}}\overline{\sigma}} -
E_{m{\mathbf{k}}+{\mathbf{q}}\sigma} + i\Gamma} \right]
e^{i{\mathbf{q}}\cdot({\mathbf{r}}_j-{\mathbf{r}}_i)+i{\mathbf{Q}}\cdot {\mathbf{r}}_j-i{\mathbf{Q}'}\cdot {\mathbf{r}}_i},
\end{eqnarray}
where ${\mathbf{Q}}$ are $N$ reciprocal lattice vectors
of the supercell lattice. The coefficients $a_{1}(u,v)$ and
$b_{1}(u,v)$ are given by
the following combinations of the coherence factors $u$ and $v$
\begin{eqnarray}
a_1(u,v)&=& v_{n{\mathbf{k}}\sigma}^*({\mathbf{r}}_i)
u_{m{\mathbf{k}}+{\mathbf{q}}\sigma}({\mathbf{r}}_i) \left(
v_{n{\mathbf{k}}\sigma}(j)u_{m{\mathbf{k}}+{\mathbf{q}}\sigma}^*({\mathbf{r}}_j)
-  u_{n{\mathbf{k}}\sigma}({\mathbf{r}}_j)
v_{m{\mathbf{k}}+{\mathbf{q}}\sigma}^*({\mathbf{r}}_j)\right),\\
b_1(u,v)&=& u_{n{\mathbf{k}}\sigma}^*({\mathbf{r}}_i)
u_{m{\mathbf{k}}+{\mathbf{q}}\overline{\sigma}}({\mathbf{r}}_i) \left(
u_{n{\mathbf{k}}\sigma}({\mathbf{r}}_j)
u_{m{\mathbf{k}}+{\mathbf{q}}\overline{\sigma}}^*({\mathbf{r}}_j) +
v_{n{\mathbf{k}}\sigma}({\mathbf{r}}_j)
v_{m{\mathbf{k}}+{\mathbf{q}}\overline{\sigma}}^*({\mathbf{r}}_j)
\right).
\end{eqnarray}
\end{widetext}
These coefficients are independent of the reciprocal vectors ${\mathbf{Q}}$ since
the eigensystem of the Hamiltonian for a given ${\mathbf{k}}$ is
invariant under shifts ${\mathbf{k}}\rightarrow{\mathbf{k}}\pm{\mathbf{Q}}$.
The remaining factors, $a_2(u,v)$ and $b_2(u,v)$, are obtained from
$a_1(u,v)$ and $b_1(u,v)$ by interchanging $u_{\sigma} \leftrightarrow
v_{\sigma}$ and $u_{\sigma}
\leftrightarrow v_{\overline{\sigma}}$, respectively. 
In Eqn. (\ref{monster}), the real-space
sum ${\mathbf{r}}_i,{\mathbf{r}}_j$, extends over a single supercell of size $N$
whereas ${\mathbf{k}}$ belongs to the reduced BZ. Below, we set
$kT=\Gamma=0.005t$. Due to the large number of diagonalizations
involved in the sum in Eqn. (\ref{monster}), we are unfortunately
unable to obtain high-resolved 2D constant energy cuts through the
BZ. However, as shown below, the
explicit $\omega$ dependence in Eqn. (\ref{monster}) allows us to
calculate e.g. ($q_x,\omega$) plots.  

Without spin order ($U=0$), the expression (\ref{monster}) reduces to the well-known result
for a homogeneous d-wave BCS
superconductor\cite{bulutnorman}. Further, when $n_h=0$ and $V=0$ but
$U \neq 0$ we find the AF state and its expected acoustic
spin cones pivoted at $(\pi,\pi)$. The spin branches are determined by
the poles of Eqn. (\ref{fullsus}) which in the homogeneous case is
given by the usual condition: $U \mbox{Re}
\chi^{+-}_0({\mathbf{q}},\omega)=1$. This condition is no longer valid
due to the $N \times N$ matrix structure
of Eqn. (\ref{fullsus}).
\begin{figure}[b]
\includegraphics[width=8.5cm]{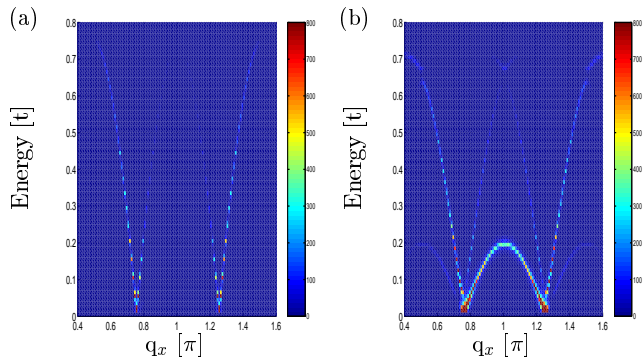}
\caption{\label{chiV0} Imaginary part of the full spin susceptibility
  $\mbox{Im} \chi^{+-}(q_x,q_y,\omega)$ at $q_y=\pi$ for $N=8$,
  $U=4.0t$, $V=2.0t$, and $t'=0.0$ (a) and $t'=-0.4t$ (b).}
\end{figure}
In the stripe phase without dSC, $(V=0)$, we show in Fig. \ref{chiV0}a the
imaginary part of the full spin susceptibility. 
Clearly, the state contains the
expected Goldstone modes shifted to $(\pi(1\pm 2/N),\pi)$\cite{batista}.
The higher harmonics (not shown) have negligible weight compared to the main
$(\pi(1\pm 2/N),\pi)$ modes. The branches broaden and lose intensity as the energy is
increased leaving very small weight near the $(\pi,\pi)$ region. At low energy, the 2D dispersion as found in constant
energy scans with $q_y \neq 0$ is dominated by circular spin cones pivoted at the IC
points. There will be four cones when the
vertical/horizontal stripe domains are equally distributed. There is
an energy range where the cones merge and form weak intensity maxima
at positions rotated $\pi/4$ from the Bragg IC points.
In the range $-0.3<t'<0.0$, the spin susceptibility is similar to
Fig. \ref{chiV0}a. However, for the more realistic range $t'<-0.3$, band-structure effects
cause a new non-Goldstone branch in the scattering response as shown
in Fig. \ref{chiV0}b. For $t'<-0.4t$, this branch
becomes fully dynamic and moves to higher energy.

The results in Fig. \ref{chiV0}
illustrate a general problem that also applies to the spin-only approaches\cite{batista}:
the outer branches, i.e. at $|q_x-\pi|>2\pi/N$, are {\sl not observed}
in the low-energy NS data\cite{cheong,niels}. This is contrary to the non-superconducting
La$_{2-x}$Sr$_x$NiO$_4$ where all four spin branches can be clearly seen in
constant energy cuts through the BZ\cite{bourges}. In the model
calculation, the intensity along the spin branches
increases monotonically with lowering the energy since the low-energy
part of the damping $\mbox{Im} \chi^{+-}_0({\mathbf{q}},\omega)_{QQ'}$
remain largely independent on the wave vector ${\mathbf{q}}$.

\begin{figure}[t]
\includegraphics[width=8.5cm]{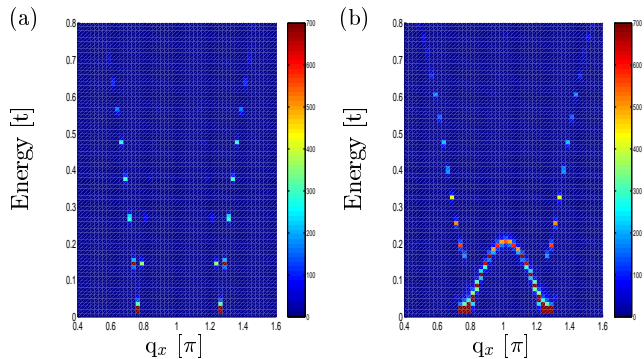}
\caption{\label{chiV}$\mbox{Im} \chi^{+-}(q_x,q_y,\omega)$ at $q_y=\pi$ for $N=8$,
  $U=4.0t$, $V=2.0t$, and $t'=-0.2t$ (a)
  and $t'=-0.37t$ (b) shown with slightly lower resolution than
  Fig. \ref{chiV0}. Note that the parameters in (b) are identical to those used in 
  Fig. \ref{configs}}
\end{figure}

What happens when superconductivity is included? Then, we expect
the matrix elements $\chi^{+-}_0({\mathbf{q}},\omega)_{QQ'}$ to strongly depend on
${\mathbf{q}}$ at low energy. Indeed, in the homogeneous dSC
phase the particle-hole continuum continues to $\omega=0$ for wave
vectors connecting the nodes of the d-wave gap. In Fig. \ref{chiV} we
show representative results for  $\mbox{Im}
\chi^{+-}({\mathbf{q}},\omega)$ in the coexisting phase.
In the low $|t'|$ regime, the result is similar to
Fig. \ref{chiV0}a. However, for parameters similar to those used in
Fig. \ref{configs}, the result is very different as shown in
Fig. \ref{chiV}b. Clearly, at low energy the inner branches are
considerably more intense that the outer. These {\sl fall within the
  particle-hole continuum and get 
strongly damped} in agreement with the NS data from Christensen {\sl
et al.}\cite{niels}. This result is one of the main points of this
paper. It is not sensitive to the size of the resulting gap $\Delta$, but rather to the specific
band-structure similar to the situation of a homogeneous dSC. Another
effect induced by $V$ is the weight around ($\pi,\pi$) which is
significantly increased in the dSC phase. Note that the overall form
of the spin fluctuation spectrum of Fig. \ref{chiV}b has the
characteristic hourglass shape\cite{tranqLBCO,araimook}. 

In the present approach we calculate the spin
response in the static stripe phase 
relevant in the underdoped regime where a pseudogap is known to exist above
T$_c$. Thus, the 'normal' state spin susceptibility in this doping
region is expected to be more like a thermally broadened version
of Fig. \ref{chiV}b as opposed to the results in Fig. \ref{chiV0}. In the optimally doped regime a spin gap
opens due to the fluctuating nature of the stripes and we expect the intensity
of the IC modes to redistribute to slightly above the
gap\cite{niels}. In the far overdoped regime the stripes presumably
disintegrate and the picture presented here eventually breaks down.

For other periodicities $N$ we find that at fixed $U$ the spin-wave velocity
is largely unchanged. Hence, the energetic
position $E_{res}$ of the $(\pi,\pi)$ 'resonance' is mainly determined by the stripe
separation. This means that $E_{res}$ increases with the doping in
the underdoped regime. This is unlike the homogeneous dSC
($U=0$, $V \neq 0$) where the intensity at $(\pi,\pi)$ {\sl can} be
increased by band-structure nesting, but where
$E_{res}$ is solely determined by
the maximum value of the dSC gap 
$\Delta$ which, contrary to experiments, decreases as the doping
increases. 

Finally, note that even though the downward mode dispersion at low
energy (Fig. \ref{chiV}b) is qualitatively similar to
that found in a pure dSC state, the
intensity distribution of the IC peaks is very
different\cite{bulutnorman}. In the former picture (mainly aimed at 
modelling YBCO and BSCCO) the resonance
intensity is maximum at $(\pi,\pi)$ and decreases with decreasing energy
until it merges with the continuum. At lower energies the response is
completely void due to the opening of a spin gap.

In summary, we have calculated the electronic structure and the
associated dynamical spin susceptibility in the stripe
phase of Eqn. (\ref{hamiltonian}). For a realistic set of parameters we find
that self-consistent solutions reproduce salient features of widely different
experimental probes including ARPES, STM and low-energy NS. This indicates that 
the stripe phase is a good starting point for describing the LSCO materials. In
the coexisting phase of IC spin, charge and dSC order, the
inner spin modes disperse toward ($\pi,\pi$) whereas the outer branches
are strongly damped. In the future, it will be
interesting to study in more detail the
dispersion of the high-energy spin fluctuations within the present approach.

This work is supported by the Danish Technical Research Council via
the Framework Programme on Superconductivity.


\begin{references} 
\bibitem{originalstripesnesting} J. Zaanan, O. Gunnarson, 
  Phys. Rev. B {\bf 40}, 7391 (1989); D. Poilblanc and T.M. Rice, {\sl
  ibid.} {\bf 39}, 9749 (1989); K. Machida, Physica (Amsterdam) {\bf 158C},
  192 (1989);  H.J. Schulz, J. Phys. (Paris), {\bf 50} 2833 (1989).  
\bibitem{tranquada} J.M. Tranquada {\sl et al.}, Nature {\bf 375}, 561 (1995). 
\bibitem{tranquadasc} J.M. Tranquada {\sl et al.}, Phys. Rev. Lett. {\bf 78}, 338 (1997). 
\bibitem{cheong} S.-W. Cheong {\sl et al.}, Phys. Rev. Lett. {\bf 67}, 1791 (1991).
\bibitem{wakimoto} H. Kimura {\sl et al.}, Phys. Rev. B {\bf 59}, 6517
(1999); S. Wakimoto {\sl et al.}, {\sl ibid.} {\bf 60}, 769
(1999); S. Wakimoto {\sl et al.}, {\sl ibid.} {\bf 63}, 172501
(2001).
\bibitem{niels} N.B. Christensen {\sl et al.}, Phys. Rev. Lett. {\bf 93}, 147002 (2004).
\bibitem{tranqLBCO} J.M. Tranquada {\sl et al.}, Nature {\bf 429}, 534 (2004). 
\bibitem{hayden} S.M. Hayden {\sl et al.}, Nature {\bf 429}, 531 (2004).
\bibitem{araimook} M. Arai {\sl et al.}, Phys. Rev. Lett. {\bf 83},
  608 (1999); H. A. Mook {\sl et al.}, {\sl ibid} {\bf 88},
  097004 (2002).
\bibitem{newtheo} M. Vojta, and T. Ulbricht, Phys. Rev. Lett. {\bf
    93}, 127002 (2004); G.S. Uhrig, K.P. Schmidt, and
  M. Gr\"{u}ninger, cond-mat/0402659; G. Seibold, and J. Lorenzana,
  cond-mat/0406589; M. Vojta, and S. Sachdev, cond-mat/0408461;
  I. Eremin {\sl et al.}, cond-mat/0409599. 
\bibitem{kaneshita} E. Kaneshita, M. Ichioka, and K. Machida,
  Phys. Rev. Lett. {\bf 88}, 115501 (2002).
\bibitem{levin} Q. Si {\sl et al.}, Phys. Rev. B
  {\bf 47}, 9055 (1993); T. Dahm, D. Manske, and
  L. Tewordt, {\sl ibid.} {\bf 58}, 12454 (1998);  A.V. Chubukov, B.
  Janko, and O. Tchernyshyov, {\sl ibid.} {\bf 63}, 180507 (2001); F. Onufrieva, and
  P. Pfeuty, {\sl ibid.} {\bf 65}, 054515 (2002); D.K. Morr, and D. Pines,
  Phys. Rev. Lett. {\bf 81}, 1086 (1998); J. Brinckmann, and P. Lee,
  {\sl ibid.} {\bf 82}, 2915 (1999).        
\bibitem{bulutnorman} N. Bulut and D.J. Scalapino,
  Phys. Rev. Lett. {\bf 47}, 3419 (1993); M. Norman,
  Phys. Rev. B {\bf 63}, 092509 (2001).   
\bibitem{batista} C.D. Batista, G. Ortiz, A.V. Balatsky, Phys. Rev. B
  {\bf 64}, 172508 (2001); F. Kr\"{u}ger and S. Scheidl, {\sl ibid} 
  {\bf 67}, 134512 (2003); E.W. Carlson, D.X. Yao, and D.K. Campbell,
  {\sl ibid} {\bf 70}, 064505 (2004).   
\bibitem{previous} Y. Chen, and C.S. Ting, Phys. Rev. Lett. {\bf 92},
  077203 (2004); J.-X. Zhu, I. Martin, and A.R. Bishop,
  Phys. Rev. Lett. {\bf 89}, 067003 (2002).
\bibitem{ichioka} M. Ichioka, and K. Machida,
  J. Phys. Soc. Jpn. {\bf 68}, 2168 (1999); M. Ichioka, and
  K. Machida, {\sl ibid.} {\bf 71}, 1836 (2002).   
\bibitem{salkola} I. Salkola, V.J. Emery, and S.A. Kivelson,
  Phys. Rev. Lett. {\bf 77}, 155 (1996); M. Granath {\sl et al.},
  Phys. Rev. B {\bf 65}, 184501 (2002). 
\bibitem{footnote1} The stripes are disordered with a flat distribution of
  average stripe distance of four lattice
  constants. Fig. \ref{configs}b is symmetrized around $k_x=k_y$
  assuming that vertical/horizontal 
  stripes contribute evenly. As discussed by Granath {\sl et
    al.}\cite{salkola} the ordered array
  already contains the essentials of Fig. \ref{configs}b
  except from the ghosty nodal weight.
\bibitem{podolsky} D. Podolsky {\sl et al.}, 
  Phys. Rev. B, {\bf 67}, 094514 (2003).
\bibitem{zhou} X.J. Zhou {\sl et al.,} Phys. Rev. Lett. {\bf 86}, 5578 (2001). 
\bibitem{howald} C. Howald {\sl et al.}, 
  Phys. Rev. B, {\bf 67}, 014533 (2003).
\bibitem{bourges} P. Bourges {\sl et al.}, Phys. Rev. Lett. {\bf 90}, 147202 (2003);
  A.T. Boothroyd {\sl et al.}, Phys. Rev. B {\bf
    67}, 100407 (2003). 
\end{references}
\end{document}